# On the Statistical Machinery of Alien Species Distribution

M G Bowler   Department of Physics, University of Oxford
             Keble Road, Oxford OX1 3RH, UK (m.bowler1@physics.ox.ac.uk)
C K Kelly    Department of Zoology, University of Oxford
             South Parks Road, Oxford OX1 3PS, UK (colleen.kelly@zoo.ox.ac.uk)

**Abstract**. Many species of plants are found in regions to which they are alien. Their global distributions are characterised by a family of exponential functions of the kind that arise in elementary statistical mechanics (a microcanonical ensemble with simple constraints; an example in ecology is MacArthur's broken stick). We show here that all these functions are quantitatively reproduced by a model containing a single parameter – some global resource partitioned at random on the two axes of species number and site number. A dynamical model generating this equilibrium is a two fold stochastic process and suggests a curious and interesting biological interpretation in terms of niche structures fluctuating with time and productivity; with sites and species highly idiosyncratic. Idiosyncrasy implies that attempts to identify *a priori* those species likely to become naturalized are unlikely to be successful.

# 1. Introduction

The study of macroecology has benefited from application of methods from the physical sciences. For example, species area relationships have been modeled mathematically [1, 2]. Species abundance distributions have been addressed with the methods of statistical mechanics [3-7]. Harte's state-variable approach to macro-ecological metrics covers both and is based on maximum entropy [8-10], very closely similar to statistical mechanics [7]. A more general review of the application of statistical mechanics in biology is given by Frank [11]. Here we apply statistical mechanics to a very different problem in macroecology, the distribution of alien species (as opposed to individuals of those species) over the globe.

In a previous study, we approached species naturalization from a global point of view, investigating the processes behind the observed distribution of 5350 naturalized species over 16 globally distributed sites. Each alien species has a footprint given by the number of sites at which it has established; the sum over all these species we termed 'the alien footprint' [12]. We determined that the observed distribution of alien species shared among sites was not an effect of geographical distance between sites, but instead indicates a statistical mechanics of naturalized species' distribution, with the alien footprint a conserved quantity. Given the diversity of sites and life history types, this finding supports the inference of complex but highly deterministic 'idiosyncratic' dynamics *sensu* Pueyo et al [4], that is, naturalization results not from any one given factor, but as a result of simultaneously being the right species, at the right place, at the right time. An additional implication of the character of the distribution is that of a 'regulator' – measured by the alien footprint - fixing the number of naturalizations in any one era.

In this paper we find an explanation unifying all the various exponential distributions in the data that are indicative of a statistical mechanics. This explanation is in terms of some global resource supporting alien establishment, divided along the two axes of site and species. We address the possible machinery for reaching these distributions in a dynamical model.

## 2. Background

In our previous paper, analysis of this substantial sample revealed certain remarkable features. The first is that the number of species found alien at $n$ sites is, for $n > 1$, exponentially distributed with $n$

$$S(n) = S_0 \exp(-\beta n) \qquad\qquad (1)$$

where the parameter $\beta$ has a value of 0.52 (see Fig 1 of [12]). Thus there are 873 species found alien to two sites (such as Wyoming and New Zealand) and 43 species alien to 8 sites and there found. No species was found at more than 13 sites. It is no surprise to find very many species at one or two sites to which they are alien and fewer at many, but an exponential distribution, as opposed to (say) a binomial or normal, must reflect some particular mechanism of community assembly. The result for $S(n)$ from our stochastic mechanism is shown in Fig.1.

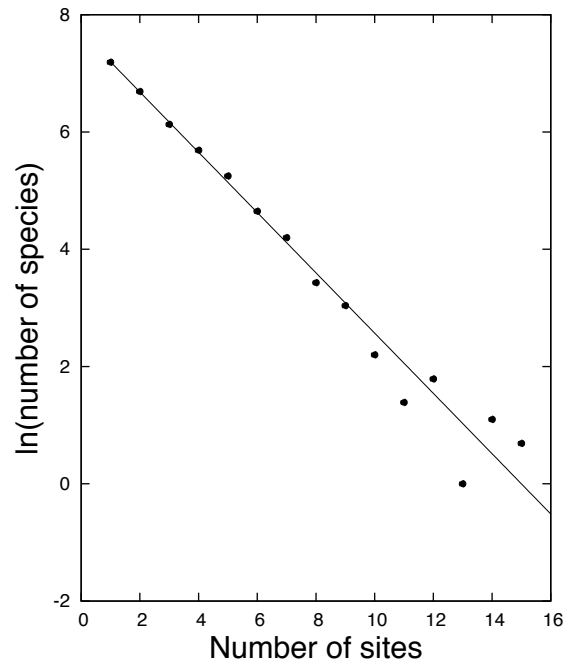

Figure 1. The model exponential distribution of the number of species over the number of sites at which they are found. The points were generated by a single run of the weighted species algorithm (section 7). The corresponding data are shown in Fig. 1 of [12].

The second remarkable observation contained in the data of our previous report is that the number of species shared pairwise has no relationship to the distance on the surface of the globe separating the sites. In Fig. 2 of that paper, each pair of sites is plotted as a point in the plane of number of species in common versus geographical separation. The distribution of points with respect to the distance between site pairs is strongly clustered geographically; the distribution of points with respect to the number of species shared falls off exponentially with that number (Fig. 3a of [12]). Yet there is no correlation between the number of naturalised species common to a pair of sites and the separation of those sites – pairs with comparatively few species in common are found in all geographical clusters and pairs with very many species in common are found at very large separations as frequently as at small separations. (The lack of any significant correlation was established quantitatively with a Mantel test.) This lack of dependence of the number of species shared on the distance separating members of the paired sites rules out any attempt to interpret the exponential falloff in the number of species at $n$ sites as an exponential attenuation with distance [12]. Thus alien species must effectively be available everywhere all the time for establishment, and dispersal may not be a significant factor. The exponential form of $S(n)$ must be associated with the availability of sites for alien species and the explanation sought in such terms.

The exponential function for $S(n)$ might suggest to the ecologist apportionment of suitable resource for alien species according to MacArthur's broken stick [13], where a fixed resource sufficient to support a specified number of alien establishments is divided at random among a fixed number of species. To the physicist that exponential suggests a version of the microcanonical ensemble in statistical mechanics, where a conserved quantity (such as energy for a simple gas) is distributed over a fixed number of agents. For this problem, the fixed resource (or conserved quantity) is the total number of alien establishments (the alien footprint of [12]) and the agents are the alien species. These two suggestions are equivalent (a third formulation is maximisation of the information entropy, with a uniform prior). This is the explanation for the exponential distribution of the number of species found at $n$ sites put forward in [12]. Thus the observations above contain two significant results: first the implication that the distribution of alien species can be understood in terms of some fixed global resource partitioned by simple random processes and secondly that dispersal seems to be universal.

The data of [12] also contain correlations with exponential distributions. These concern pairs and triplets of sites and the number of alien species that they share – for example, Wyoming and New Zealand have in common 164 species alien to both. The number of site pairs sharing $p$ species is exponentially distributed in $p$. Similarly the number of triplets of sites sharing $t$ alien species is exponential in $t$ (and the pattern repeats even for quartets). These correlations depend not only on the distribution of species over sites but also on the distribution of sites over species. There are 560 triplets of sites and 120 pairs; their exponential distributions are well defined. The distribution of the number of sites $M(s)$ at which $s$ species are found cannot, for a single sample of only 16 sites, look like a continuous distribution. An ideogram (Fig.2, *lower panel*) shows individual site occupancy widely spread, but tending to cluster at low values; these data are consistent with having been drawn from an underlying exponential distribution [12]. That the origin of these additional exponential distributions is related to that of the exponential $S(n)$ was a matter of conjecture. We have now shown that if the same global resource divided randomly over species is also divided randomly over the sites axis, then all these features are unified within a simple statistical model. We address the possible machinery for reaching these distributions in a dynamical model, which turns out to have aspects not purely local.

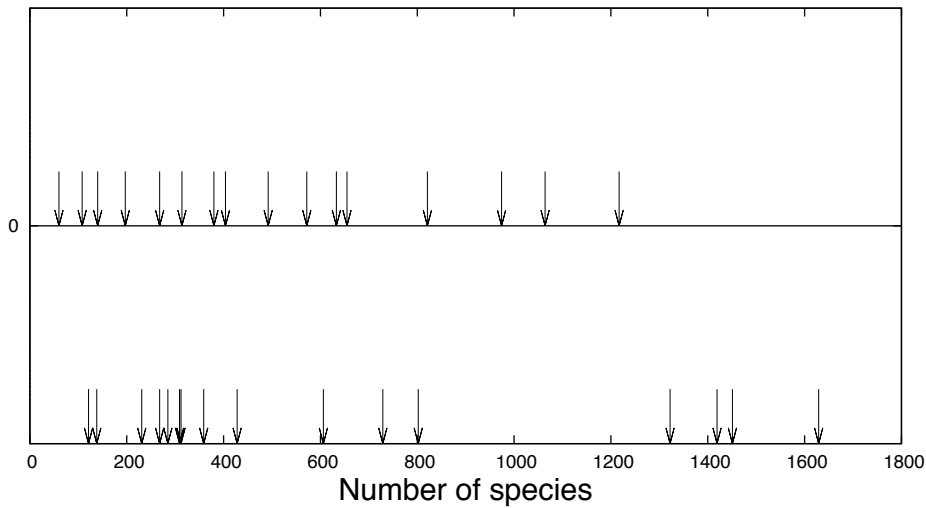

Fig.2 The arrows indicate the number of species at each of 16 sites, covering a range from ~ 100 to ~ 1600. The *upper panel* is a stochastic model simulation for an exponential probability distribution function, the *lower panel* displays the data of [12]Kelly et al, not illustrated in that paper.

## 3. Indicators of biological machinery

Having conjectured that the combination of an exponential distribution $S(n)$ and an exponential singlet distribution $M(s)$ might generate the exponentially distributed multiplets, we considered processes, in the context of statistical mechanics, that result in these two exponentials and how they might be combined. In order to explore correlations, we need an algorithm yielding the contents of every cell in a matrix $N(J,K)$ in which rows represent the 16 sites (from Chile to Wyoming) and columns the thousands of alien species involved (species 5350 is *Zygophyllum fabago*). From such a matrix the pair and other multiplet distributions can be constructed. An algorithm containing the observed distributions must exist, but Nature's algorithms are not always easy to find and interpret. Two aspects of the data led to construction of a dynamical niche based picture and an algorithm which yields a species-site population matrix in excellent agreement in all respects with the data set. The first aspect is the complete lack of dependence of the number of species shared on the separation over the globe of the sites in a pair. The second is the possibility of interpreting the exponential distributions $S(n)$ and $M(s)$ in terms of analogues of niches opening and closing, accepting and rejecting alien species.

If the system is dynamical (like the internal workings of a gas) as suggested by the fact of the spread of alien species over the globe, the dynamics of approach to the equilibrium distribution can be modelled with a simple master equation (see e.g. [6, 7, 14])

$$\frac{dS(n)}{dt} = -(r_n^+ + r_n^-)S(n) + r_{n-1}^+ S(n-1) + r_{n+1}^- S(n+1) \qquad (2)$$

Here, $S(n)$ is the number of species found at $n$ sites. The parameters $r_n^-$ and $r_n^+$ are rates at which a species vanishes from some site or appears at a site from which it was previously absent. The content of Eq.(2) is that if a species is present at $n$ sites $S(n)$ is reduced by 1 if that species vanishes from one site or if that species appears at one additional site. Similarly, $S(n)$ is augmented by 1 by adding to a new site a species present at $n$-1 sites, or by losing from a site a species present at $n$+1 sites. The equation evolves to a steady state at which

$$S(n+1) = \frac{r_n^+}{r_{n+1}^-} S(n)$$

Iterating this equation, the solution is an exponential in $n$ provided that the ratio of rates $r_n^+ / r_{n+1}^-$ is independent of the number of sites $n$ at which the species is present. Then we can write $r_n^+ = r^+ f(n+1)$, $r_n^- = r^- f(n)$ and obtain

$$S(n) = S(0) \left( \frac{r^+}{r^-} \right)^n \qquad (3)$$

a negative exponential if $r^+ < r^-$. If a species is at $n$ sites, it is removed from one of those sites a little more often than it is added to a new site. In (3), $S(0)$ is the number of species present at no alien site at any given time, after a dynamic equilibrium has been reached (in which the individual species wander stochastically in the number of sites at which they are found). The most economical way of achieving the exponential solution (3) is to have neither of the parameters $r_n^+$ and $r_n^-$ depend on $n$, $f(n) = 1$. We have made this assumption for the purpose of simulating Eq.(2) as a stochastic process to fill the matrix $N(J,K)$. Any algorithm leading to (3) will of course generate an exponential distribution (see section 8a).

## 4. Implementation as a stochastic process

The content of Eqs. (2) and (3) can be simulated very simply for the purpose of obtaining the distribution of species over sites, represented by the matrix $N(J,K)$ introduced in section 3. If species $J$ is present at site $K$ then that element of the matrix is 1, otherwise 0. Apply the following operations to this matrix- it is perfectly reasonable to start with all elements zero. Choose at random a species $J$. Then choose at random whether to open a new site for this species or remove one of the sites already filled, the ratio of choices being $r^+ / r^-$. If the choice is for putting $J$ into a site it does not already occupy, choose one of the empty sites and change that element of the matrix from 0 to 1. If on the other hand the lot was cast for emptying a site, choose one of the elements for $J$ with occupancy 1 and change it to 0. For the purpose of generating $S(n)$ it does not matter by what recipe the empty site to be filled is chosen, nor the full site emptied. Repeat this operation a sufficiently large number of times for the equilibrium configuration to emerge (further repetition changes which species are at a given $n$ sites but leaves the distribution unchanged). Then for each species count the number of filled $K$ elements (sites) and count that species into the appropriate bin to yield the distribution $S(n)$. It is an exponential and the exponent parameter $\beta$ of (1) is given by

$$\beta = -\ln\left(\frac{r^+}{r^-}\right) \tag{4}$$

Thus this picture of alien species available for suitable niches, which open and close at rates independent of the total niche space already filled, straightforwardly accounts for the exponential distribution of the number of species as a function of the number of sites (as does a single multiple fracture of MacArthur's stick; [13]. The results of a model calculation are shown in Fig.1, which may be compared with Fig.1 of [12]. Simulation was not needed to obtain this result, but this is only the $S(n)$ part of the problem; there is the second axis concerning $M(s)$ and simulation is desirable to generate $N(J,K)$ and study correlations, statistical fluctuations and the time evolution of the system.

## 5. Complications in two dimensions

Table 1 shows a small portion of the matrix $N(J,K)$ from which the model results in Figs. $1-3$ of this paper were drawn.

```
0 0 0 0 0 0 0 0 0 0 0 0 0 0 0 0
0 0 0 0 0 0 0 0 0 0 0 0 0 0 0 0
1 1 1 1 0 0 1 0 0 0 0 1 0 0 0 0
0 0 0 1 0 0 0 0 0 0 0 0 0 0 0 0
0 0 0 0 0 0 0 0 0 0 0 0 0 0 0 0
1 1 1 0 0 1 0 0 1 0 0 0 0 0 0 0
0 0 0 0 0 0 0 0 0 0 0 0 0 0 0 0
0 0 0 0 0 0 0 0 0 0 0 0 0 0 0 0
0 0 0 0 0 0 0 0 0 0 0 0 0 0 0 0
0 0 0 0 0 0 0 0 0 0 0 0 0 0 0 0
0 0 0 0 0 0 0 0 0 0 0 0 0 0 0 0
1 1 0 0 0 0 0 0 1 0 1 0 0 0 0 0
0 0 0 0 0 0 0 0 0 0 0 0 0 0 0 0
0 0 0 0 0 0 1 1 1 0 0 0 0 0 0 0
0 1 1 0 0 1 0 0 1 1 0 0 1 0 0 0
0 0 1 0 0 0 0 0 0 0 0 0 1 0 0 0
0 0 0 0 0 0 0 0 0 0 0 0 0 0 0 0
0 0 0 0 0 1 0 0 0 0 0 0 0 0 0 0
0 0 0 0 0 0 0 0 0 0 0 0 0 0 0 0
0 0 0 0 0 0 1 0 0 0 0 0 0 0 0 0
0 0 0 0 0 0 0 1 0 0 0 0 1 0 0 0
0 0 0 0 0 0 0 0 0 0 0 0 0 0 0 0
1 0 1 0 0 0 0 1 0 0 0 0 0 0 0 0
1 1 0 1 0 0 0 1 1 0 1 0 0 0 1 0
1 1 0 1 1 1 1 1 0 0 1 0 0 0 1 0
0 0 0 0 0 0 0 0 0 0 0 0 0 0 0 0
0 0 0 0 0 0 0 0 0 0 0 0 0 0 0 0
```

Table 1. A portion of the model calculation of $N(J,K)$. Each row gives the status of a single species at each of 16 sites. In the full matrix there are thousands of species and hence thousands of rows.

The number of sites at which a species is present is found by counting horizontally along the appropriate row; thus the third species in the list is found at six sites and the fourth is found at one. The number of species found at any given site is obtained by counting vertically. It is immediately clear that although an exponential distribution *S(n)* results from opening or closing as appropriate any of the sites for a given species (a species algorithm) regardless of how selected, the distribution of species over sites is determined by how that selection is made. Thus if a choice is made at random among the filled sites as to which to delete and similarly a random choice is made of which empty site to fill, then all sites are being treated in the same way and the distribution of occupancy will be approximately normal about the mean (approximately 600 species per site). This is very unlike the data where the singlets are consistent with being drawn from an exponential distribution, extending from under 200 to over 1600 species at a site (Fig. 2). Thus in a species algorithm, selection of the next site to open or close must be made according to a recipe that will yield a singlet distribution consistent with exponential.

## 6. Distribution of site species populations

If the matrix *N(J,K)* is addressed differently, there is an obvious way of generating an exponential singlet distribution. Choose at random any site (out of only 16) and then either add a species not already present or, slightly more often, delete a species at that site. Regardless of how a species is chosen, an exponential distribution (more accurately, a set consistent with having been drawn from an exponential distribution) will result from this site algorithm. However, in looking for both an exponential distribution of *S(n)* and simultaneously an exponential distribution of singlets *M(s)* with occupation number *s*, elements of *N(J,K)* must be changed consistently, working both horizontally (species algorithm) and vertically (site algorithm). For the horizontal approach yielding an exponential *S(n)* the ratio $r^+/r^-$ for species gaining or losing sites needed for an exponential matching the data is ~ 0.6, yet to generate a singlet distribution with a mean of ~ 600 (as observed) requires a ratio $r^+/r^-$ for sites gaining or losing species ~ 0.998. The length of the stick to be broken, that is, the total resource to be partitioned, is in both cases the sum of elements in the matrix, the alien footprint. Nonetheless, the exponential singlets distribution can be made consistent with the mean number of sites per species and yet be attributed to opening and closing of niche structures in essentially the same way as *S(n)*.

We envisage a site as having a degree of receptivity to alien species (rather than specific niches) and that receptivity fluctuating with time. It might correspond to capacity for a certain number of alien species, that number increasing or decreasing by amounts independent of the number itself. This could be described by a master equation for *M(s)* of the same type as (2) and with a suitable ratio of the frequency of increasing to decreasing capacity yields outputs consistent with being drawn from an exponential with a mean ~ 600. While site populations from a single run of the site algorithm can only be displayed as an ideogram, the sum over many separate independent runs turns into a histogram of clearly exponential nature, realising the underlying probability distribution. We suppose, then, that alien species are available to colonize sites whose receptivity has evolved to an exponential probability distribution. The recipe for picking sites from which to remove a species or add a new species must reflect this underlying receptivity, expressed in the form of weights.

One way is to generate a set of individual occupations with a single run of a site algorithm. Examples of two independent runs are in Table 2, where the left hand column specifies the site $K$ and the right hand columns the number of potential species accommodated. These sets of numbers give the relative receptivity of each site at the time the evolution was sampled. The label $K$ has no significance, but ordering the sites by receptivity creates a rank $\{R\}$. Separate runs give sets of numbers drawn from the same underlying exponential distribution. If all that is wanted from the site algorithm is a set of numbers of species at each site, the individual species can be ignored and the site algorithm collapsed to the one dimension $K$, computationally much more efficient. Table 2 indicates how much scatter there is between runs; we have chosen rather to present results obtained using averaged weights determined by the underlying exponential probability distribution.

| 1 | 97 | {15} | 256 | {10} |
|---|---|---|---|---|
| 2 | 782 | { 6} | 1165 | { 3} |
| 3 | 468 | { 8} | 435 | { 5} |
| 4 | 1526 | { 3} | 349 | { 7} |
| 5 | 133 | {13} | 873 | { 4} |
| 6 | 859 | { 5} | 427 | { 6} |
| 7 | 1636 | { 1} | 163 | {13} |
| 8 | 187 | {12} | 2874 | { 1} |
| 9 | 767 | { 7} | 1781 | { 2} |
| 10 | 1136 | { 4} | 168 | {12} |
| 11 | 75 | {16} | 220 | {11} |
| 12 | 1601 | { 2} | 307 | { 9} |
| 13 | 106 | {14} | 16 | {16} |
| 14 | 467 | { 9} | 328 | { 8} |
| 15 | 242 | {11} | 40 | {14} |
| 16 | 357 | {10} | 35 | {15} |

Table 2. Results from two independent runs of the site algorithm. {R} denotes site rank.

## 7. A method of averaged weighting

If the site algorithm numbers are ordered by rank, the averages of $s_R$ over many independent runs of the stochastic algorithm can be calculated for each $R$. Thus for the most receptive sites $s_1 \sim 2000$ but for the least $s_{16} \sim 50$ (see Table 2). As samples accumulate the means cluster ever closer to a straight line with $s_R$ proportional to $\ln R_0 - \ln R$, where the constant $R_0$ emerging from the accumulated runs is close to 17. An exact calculation of these averages can be made analytically from the underlying exponential probability distribution and shows that the above simple relation is sufficiently accurate. The relative weight (or receptivity) of a site of rank $R$ we therefore took as $\ln R_0 - \ln R$. These averaged rank ordered receptivities were used in weighting the rank ordered sites when applying the species algorithm to produce our final results shown in the figures.

To implement the species algorithm with the sites appropriately weighted is straightforward. If a chosen species is to be removed, choose one of its occupied sites at random and delete the species. If on the other hand a species is to be added to a site, choose among the (ordered) sites without this species according to their relative weights (receptivities) $\ln R_0 - \ln R$. Thus the site with $R = 15$ seldom gains a new species. This implements the notion of some sites having wandered over aeons to a state more receptive of alien species than others. Fig. 2 shows ideograms of the singlet populations for both the data and the output of our final algorithm. They are consistent with each other. Since a tick mark is made for each individual population number, the exponential nature of the underlying distributions is manifest in ticks being denser at the low end of the occupancy axis. Similar distributions are obtained for separate runs of the singlet algorithm (for example, ideograms drawn from Table 2) and many runs of the algorithm for singlet populations can be summed into histogram bins and accumulated to define an exponential.

When the species algorithm is run according to this weighted recipe the output of a single run is in admirable agreement with the data, from which the value of the alien footprint for $n > 1$ and the 2049 such species determines the ratio $r^+/r^-$ (0.6, the only parameter fine tuned) for the species algorithm. The simulation then yields a distribution $S(n)$ with a mean of about 2 sites per species (Fig. 1). The same number also determines for the 16 sites the value of $r^+/r^-$ for the site algorithm, with a mean of $\sim 600$ species per site (Fig. 2). The full matrix $N(J,K)$ thus generated was interrogated to yield the number of site pairs sharing species as a function of the number of species shared and the number of triplets as a function of the number of species common to three sites at which they are alien. The simulation reveals that these shapes are indeed exponential; Fig. 3 does not differ significantly from the pairs and triplets found in the data, shown in Fig. 3 of [12].

The results from our simulation shown are for 5700 species loose and drifting in and out of naturalisation, to match the results of [12] containing approximately 3400 species under the exponential in Fig. 1 of that paper. The distribution $S(n)$ is exponential by construction; the pair and triplet distributions emerge as exponential. The parameters of these exponentials are then given; the curves shown in Fig.3 are not fits to the data, but were calculated from the distribution $S(n)$, just as in [12]. These conclusions are robust; see section 8 below.

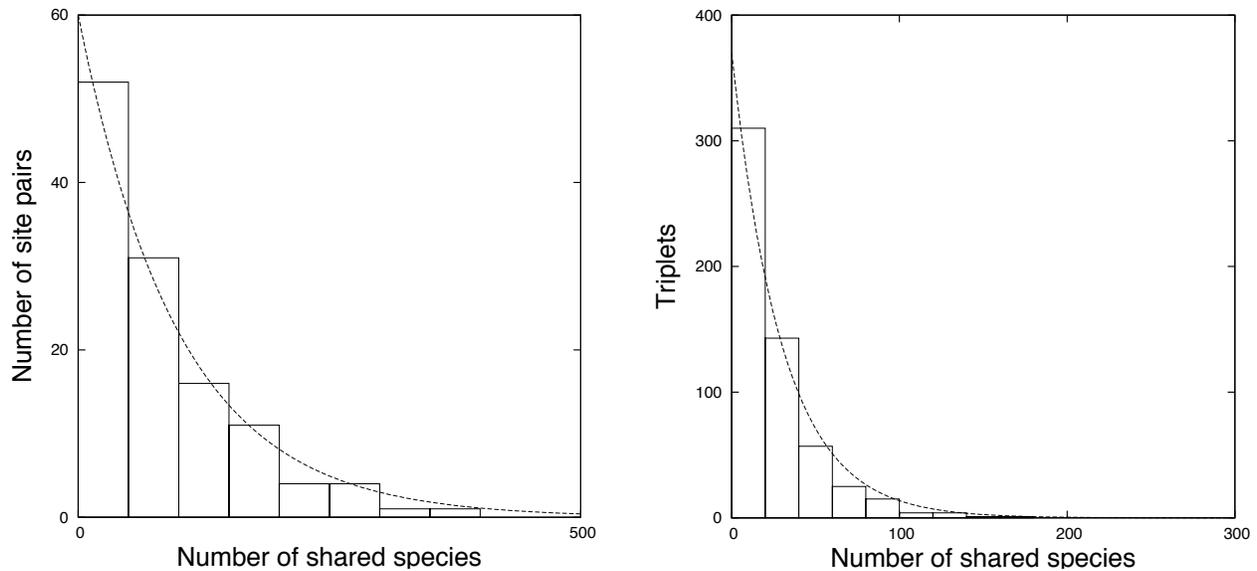

Fig.3 Modelled distributions for multiplets. *Left panel*: The number of site pairs as a function of the number of shared species. *Right panel*: The number of triplets as a function of the number of species shared. Both are exponential, generated by the weighted species algorithm. To be compared with the data shown in Fig.3 of [12].

## 8. Other ways of filling the matrix

### *8a. Variations on the dynamical algorithm*

Forming the weights from a single run of the site algorithm shows the same features as described in section **7**, but the output is noisier. It is not known what single run of the site algorithm might be most representative of the real world; nonetheless, if relative weights are taken from, for example, either of the runs listed in Table 2, the resulting multiplet distributions are in good accord with Fig.3 and with the observations. The alien footprint for the successfully naturalised species and 16 sites is the single parameter in the model, unifying the various exponential distributions found by [12]. We finally note that in the above discussion we have tacitly assumed that the distribution of site receptivities has settled down before the alien species are unleashed. While this seems a natural assumption to make, it is not necessary and our conceptual structure is more general. The site algorithm (operating in the $K$ dimension) can be embedded in the weighted species algorithm and (as an example) a site capacity updated each time a species has its complement of sites updated. The equilibrium distributions nonetheless emerge in agreement with Figs. 1-3 (but noisier).

These algorithms generated directly an exponential $S(n)$ and distributions $M(s)$ drawn from an underlying exponential probability distribution. For convenience the probabilities of adding a new site to a species or removing a site from the species are independent of the number of sites at which that species is present; similarly for augmenting or decreasing a site's complement of species. It is clear that any choice of the function $f(n)$ in implementing the master equation (2) would be equally successful in matching the data; all that is required is that $r_n^+ = r^+ f(n+1)$ with $r_n^- = r^- f(n)$.

### 8b. A static algorithm

The matrix $N(J,K)$ can be populated assuming the underlying exponential probability distributions without considering the machinery by which they are reached. The matrix $N(J,K)$ is then populated by first calculating analytically the most probable site receptivities as a function of rank $R$. The number of species alien to only one site is – say – 1300, at two sites 800, at three sites 450 and so on, falling exponentially with the number of sites. In such a scheme species numbers 1 – 1300 are assigned each to a single site randomly in accord with the relative receptivities. The next 800 are assigned to two sites, in accord with the relative receptivities and so on. The matrix $N(J,K)$ thus constructed can be interrogated to find the number of pairs of sites as a function of the number of species they have in common, similarly for the number of triplets. When this static algorithm is implemented, the version of Fig.3 produced is in excellent agreement with that from the dynamical algorithm.

## 9. Discussion

The naturalisation of many thousands of plant species over the globe is inherently a dynamical process of great complexity, yet has emergent properties described by a simple statistical model. A resource (represented by the alien footprint $\sum_n nS(n)$ and global in nature) is partitioned at random among 16 sites and then again at random among over $\sim 5000$ species alien to those sites. All sites are treated as equivalent and all species likewise. The most probable configuration is then drawn from exponential probabilities for the number of species found at $n$ sites and the number of sites containing $s$ species. These distributions are a simple application of the microcanonical ensemble in statistical mechanics, or of maximum entropy with uniform priors. In the language of ecology, this is generating distributions of alien species according to a twofold example of MacArthur's broken stick. MacArthur's broken stick model failed in its original application to the very different problem of species abundance, but it is realised in abundance of species - the global distribution of alien species. Any algorithm that generates an exponential distribution of species over the number of sites at which they are present and simultaneously species populations of sites drawn from an exponential distribution yields exponential distributions for pairs and triplets of sites, as shown in Fig.3 above (*c.f.* Fig.3 of [12].

If these distributions are the result of dynamical processes represented by the master equation (2) then the exponential distributions imply curious and interesting conditions. The ratio formed from the rate for introducing a species, present at $n$ sites, to an unoccupied site divided by the rate of losing that species, when present at $n+1$ sites, from an occupied site is independent of $n$. There is an analogous condition for assigning species to sites. As set out in section 3, the requirement can be represented algebraically as $r_n^+ = r^+ f(n+1)$, $r_n^- = r^- f(n)$. It was convenient to choose $f(n) = 1$ which satisfies this condition trivially, but the rates at which species become naturalised or extinct at any site depend on the processes involved. The appropriate function $f(n)$ is not suggested by the data nor determined by the model, but there are biological implications, usefully illustrated by two rather different examples

The first is for $f(n) = 1$. Then the rate at which a species gains a new site is independent of the number it already occupies, which does not seem implausible because occupied sites are not expected to reproduce and give rise each to new ones. However, the rate at which a species loses a site is also independent of the number at which it is present, yet there is an intuitive expectation that the more sites at which a species is present the faster it will vanish from one of them. This is reinforced by the notion of *per capita* death rates for individuals in populations of given species, but the ecology of alien establishments is a very different problem. The intuitive expectation is based upon an assumption, usually not made explicit, that events such as extinction of a particular species have the same probability of occurring at any site and are uncorrelated. It does not have to be so. The requirement could be met if at any particular time a species is vulnerable at only one site and it is at that site that it takes the hit. The ecological significance would be that establishment and extinction of alien species are not determined wholly by independent local processes; the global picture is important. For this case of $f(n) = 1$ it could be thought of in terms of some global niche space for a given species, that space expanding and contracting by absolute amounts (as opposed to fractional changes). Naturalisation is not, in the application of the master equation, purely a local process.

The second example applies to the case where extinctions are random and uncorrelated. In this case the function $f(n) = n$ and this does agree with intuitive notions. If so, then the observation of the exponentials in the distribution of alien species requires that the rate at which a species present at $n$ sites is naturalised at one more is proportional to $n + 1$. The most obvious interpretation would be that colonies give rise to new colonies at the same rate as the parent site. This is not the only possible interpretation, but other possibilities also seem to depend on fine tuning. In no case is there any implication that a species currently found at many sites is likely to be a menace at any of these sites or at any site at which it is not currently naturalised. The feature common to all these scenarios generating exponential distributions is a global ceiling.

## 10. Conclusions

This work has established a simple theoretical foundation unifying the separate observations to be found in [12], much as conjectured therein. The inferences about the biology of alien species and sites in that work are thus strongly supported.

Perhaps the most important inference comes from the mere applicability of elementary statistical mechanics to this problem. In these models, no site is special and it is as a result of random processes that it reaches a particular capacity for alien species. Similarly, no species is special and it reaches a particular number of sites through random processes. These equivalences among sites and among species might come about through genuine identity, known as neutrality in the context of species abundance distributions, but given the variety of sites and species, this is not credible. Rather, the equivalence must arise through extreme individuality of both sites and species – the idiosyncrasy of [3, 4]. This idiosyncrasy may be highly relevant to the general problem of community assembly – [12] describe two documented examples of species within guilds distributed exponentially over sites [15, 16].

We have an indication of the machinery underlying the statistical mechanics of alien species - a global reservoir of alien species, each awaiting the opening or rejected by the closing of suitable geographic sites. In the context of such a dynamical model, the process of naturalisation is not purely local; these are global aspects. Underlying the ebb and flow of species is a global conserved quantity, the number of alien establishments – the alien footprint of [12]. This alien footprint, for a given number of sites and of species, is (mathematically) the only free parameter in the model. Biologically, it represents some resource that may be related to global net primary productivity and it is an instance of the process of 'biotic resistance' and the fundamental regulation of community diversity [17-22]. $CO_2$ has been shown elsewhere to enhance net primary productivity above the general action of climate change (temperature and water availability; [23] suggesting that it may be the fundamental factor responsible for the ongoing rate of species naturalizations currently being observed throughout the globe [24-28].

In the early stages of this research CKK was supported in part by DEB 0713866 and a grant from the National Geographic Society.